\documentclass[conference]{IEEEtran}
\IEEEoverridecommandlockouts
\usepackage{cite}
\usepackage{amsmath,amssymb,amsfonts}
\usepackage{algorithmic}
\usepackage{graphicx}
\usepackage{textcomp}
\usepackage{xcolor}
\usepackage{url}
\usepackage{comment}
\usepackage{csquotes}

\usepackage{tabularx}
\usepackage{multirow}
\usepackage{colortbl}

\usepackage{booktabs}

\usepackage[acronym]{glossaries}
\newacronym{acta}{ACTA}{Applied Cognitive Task Analysis}
\newacronym{cta}{CTA}{Cognitive Task Analysis}
\newacronym{emt}{EMT}{emergency medical technician}
\newacronym{hta}{HTA}{Hierarchical Task Analysis}
\newacronym{sme}{SME}{subject matter expert}
\newacronym{uhh}{UHH}{Universit\"at Hamburg}
\newacronym{uke}{UKE}{University Medical Center Hamburg-Eppendorf}

\definecolor{maroon}{cmyk}{0,0.87,0.68,0.32}

\def\BibTeX{{\rm B\kern-.05em{\sc i\kern-.025em b}\kern-.08em
    T\kern-.1667em\lower.7ex\hbox{E}\kern-.125emX}}

\usepackage{xspace}
\newcommand{\stageONE}{Task Diagram Interview\xspace}
\newcommand{\stageTWO}{Knowledge Audit\xspace}
\newcommand{\stageTHREE}{Simulation Interview \& Workshop\xspace}
\newcommand{\stageFOUR}{Mockup Refinement\xspace}

\usepackage{tikz}
\usepackage[framemethod=TikZ]{mdframed}
\newcounter{theo}
\renewcommand{\thetheo}{\arabic{theo}}
\newenvironment{theo}[2][]{
\refstepcounter{theo}
\ifstrempty{#1}
{\mdfsetup{
frametitle={
\tikz[baseline=(current bounding box.east),outer sep=0pt]
\node[anchor=east,rectangle,fill=gray!20]
{\strut Lesson Learned~\thetheo};}}
}
{\mdfsetup{
frametitle={
\tikz[baseline=(current bounding box.east),outer sep=0pt]
\node[anchor=east,rectangle,fill=gray!20]
{\strut Lesson Learned~\thetheo:~#1};}}
}
\mdfsetup{innertopmargin=5pt,innerbottommargin=10pt,linecolor=gray!20,
linewidth=2pt,topline=true,%
frametitleaboveskip=\dimexpr-\ht\strutbox\relax
}
\begin{mdframed}[]
\label{#2}}{\end{mdframed}}

\begin{document}

\title{Lessons Learned from Customizing and Applying ACTA to Design a Novel Device for Emergency Medical Care}

\author{
    \IEEEauthorblockN{
        Christoph Stanik\IEEEauthorrefmark{1}, 
        Tim Puhlf\"urß\IEEEauthorrefmark{1}, 
        Anne Mahler\IEEEauthorrefmark{2}, 
        Phillip Brenya Sasu\IEEEauthorrefmark{2}, 
        Wikhart Reip\IEEEauthorrefmark{2}, and 
        Walid Maalej\IEEEauthorrefmark{1}
    }
    \IEEEauthorblockA{
        \IEEEauthorrefmark{1}
        \textit{Universit\"at Hamburg} \\
        Hamburg, Germany \\
        E-Mail: christoph.stanik@uni-hamburg.de, tim.puhlfuerss@uni-hamburg.de, walid.maalej@uni-hamburg.de
    }
    \IEEEauthorblockA{
        \IEEEauthorrefmark{2}
        \textit{University Medical Center Hamburg-Eppendorf}\\
        Hamburg, Germany \\
        E-Mail: an.mahler@uke.de, p.sasu@uke.de, w.reip@uke.de
    }
}

\maketitle

\begin{abstract}
Preclinical patient care is both mentally and physically challenging and exhausting for emergency teams. 
The teams intensively use medical technology to help the patient on site. 
However, they must carry and handle multiple heavy medical devices such as a monitor for the patient's vital signs, a ventilator to support an unconscious patient, and a resuscitation device. 
In an industry project, we aim at developing a combined device that lowers the emergency teams' mental and physical load caused by multiple screens, devices, and their high weight. 
The focus of this paper is to describe our ideation and requirements elicitation process regarding the user interface design of the combined device. 
For one year, we applied a fully digital customized version of the \gls{acta} method to systematically elicit the requirements.
Domain and requirements engineering experts created a detailed hierarchical task diagram of an extensive emergency scenario, conducted eleven interviews with \glspl{sme}, and executed two design workshops, which led to 34 sketches and three mockups of the combined device's user interface.
Cross-functional teams accompanied the entire process and brought together expertise in preclinical patient care, requirements engineering, and medical product development. 
We report on the lessons learned for each of the four consecutive stages of our customized \gls{acta} process.
\end{abstract}

\begin{IEEEkeywords}
requirements elicitation,
digital workshop,
applied cognitive task analysis,
lessons learned
\end{IEEEkeywords}

\section{Introduction} \label{sec:introduction}

\textbf{Context.}
Preclinical patient care has reached a very high standard in many parts of Europe. 
This standard is associated with high education and training, and the intensive use of medical technology directly at the emergency scene.
To adequately care for a critically ill patient, the emergency team must carry heavy equipment, of which about 20 kg belong to the defibrillation, pacing, monitoring, and ventilation, including the oxygen bottle.

\textbf{Problem.}
Although medical equipment became more compact and lighter in recent years, the number of devices and the total weight has increased. 
This puts a lot of physical stress on the team members.
It is cumbersome to carry every device to the emergency scene as it requires multiple team members to carry them.
Furthermore, indoor emergency scenes, like small apartments, offer very limited space to place the equipment.
Therefore, depending on the scene and the communicated state of the patient, the emergency team decides beforehand which devices they carry from the ambulance to the scene.
In case they need additional equipment, one of the team members has to go back to the ambulance to bring the missing devices.
The situation becomes even more challenging if the emergency team has to transport the patient from the scene to the ambulance.
During the transport, the team has to take care of the patient's well-being while not accidentally pulling any cable, catheter, or the oxygen tubing, which might endanger the patient and triggers loud alarms.
Also, the team must be able to interact with the devices during transport to take care of the patient and to adjust parameters if needed.

\textbf{Solution.}
The overall goal of this project is to design and develop a concept of a medical product that integrates the features of multiple devices that are applied during preclinical patient care.
Emergency teams shall have to carry less and lighter equipment, without missing necessary functionality.
This paper presents our fully digital ideation and requirements elicitation process for the envisioned graphical user interface of the combined medical product concept.
The solution results from a joint project of two industry partners, and the scientific partners the Department of Anesthesiology of the \gls{uke}, and the Informatics department of the \gls{uhh}.

\textbf{Method.}
As preclinical patient care requires a strong cognitive focus, we applied a variant of the \gls{acta} method \cite{militello:ergonomics:1998}, which aims to make task analysis techniques more accessible to practitioners who are not experts in cognitive psychology.
In this paper, we report on the four stages that we conducted during this project: \textit{\stageONE}, \textit{\stageTWO}, \textit{\stageTHREE}, and \textit{\stageFOUR}.
We designed each stage around empirical methods like semi-structured interviews and iterative design workshops.
Besides reporting on how we applied \gls{acta} digitally, we further highlight our lessons learned for practitioners interested in following a similar approach.
The process entails many design challenges, particularly accessibility aspects like displaying only the information that is required in the current therapy context to avoid overwhelming the emergency team.

\textbf{Results.} 
We created one hierarchical task diagram of an extensive emergency scenario, conducted eleven interviews with \glspl{sme}, and executed two design workshops, which led to 34 sketches and three mockups of the combined device's user interface.
Additionally, we extracted six lessons learned from our process.

\textbf{Structure.}
In Section \ref{sec:related_work}, we discuss the related work regarding the \gls{acta} method, user interface design in the medical domain, and designing new devices during online workshops.
Section \ref{sec:project} gives an overview of our \gls{acta} approach.
Afterward, we describe each stage of our process in detail, starting at Section \ref{sec:stage_1}, which is about the \textit{\stageONE}. 
Section \ref{sec:stage_2} presents the \textit{\stageTWO} stage. Section \ref{sec:stage_3} explains the stage \textit{\stageTHREE}. 
Section \ref{sec:stage_4} details the stage \textit{\stageFOUR}.
Finally, we discuss the applicability of \gls{acta} in this project and our lessons learned in Section \ref{sec:discussion} and conclude our work in Section \ref{sec:conclusion}.
\section{Related Work} \label{sec:related_work}

\subsection{Applied Cognitive Task Analysis}

Militello and Hutton designed the \acrlong{acta} method as a streamlined version of the \glsfirst{cta} \cite{militello:ergonomics:1998}. 
The researchers aimed to make the task analysis techniques more accessible to practitioners who are not experts in cognitive psychology but require an analysis about the specific difficulties in tasks. 
\Gls{acta} consists of four stages: the development of a \textit{task diagram}, the \textit{knowledge audits}, the \textit{simulation interviews}, and the summarization of all relevant difficulties in a \textit{cognitive demand table}.
In an evaluation study, the researchers demonstrate that (1) \gls{acta} is easy to use, (2) the provided guide for the knowledge audits aids to be flexible, (3) the output of the interviews is clear, and (4) the representation of the collected cognitive demands is a useful basis for further proceedings. 
Nevertheless, they criticize the lack of well-established metrics for testing the reliability and validity of \gls{acta}. 
Moreover, they assume that \gls{acta} gathers less comprehensive and more superficial information when compared to more systematic techniques like the critical decision method \cite{klein:IEEE:1989} or the cognitive analysis \cite{rasmussen:elsevier:1986}.
While we found the critics to be true in our \stageTHREE stage, we could add meaningful details and create more sophisticated requirements in the \stageFOUR stage. 

\subsection{User Interface Design in the Medical Domain}

Pickup et al. \cite{pickup:JMIR:2019} used \gls{acta} in combination with a user-centered design approach to generate user interface specifications of a novel device that facilitates the resuscitation process for a newborn child.
However, they modified the \textit{Simulation Interview} stage of \gls{acta} to an interactive scenario-based workshop with twelve \glspl{sme} (pediatric doctors, neonatal nurses, and practitioners, midwives). 
Besides evaluating the experts' cognitive difficulties during the neonatal resuscitation process, the researchers aimed to design a mockup of the novel device's user interface during the workshop.
Due to its similarity in the domain and overall goal, this project inspired the structure of our study.
However, since the device is only applicable in less diverse scenarios, interviews with two pediatric doctors were sufficient for Pickup et al. to collect a sufficient amount of information about cognitive difficulties in neonatal resuscitation. 
In contrast, we had to consider various possible scenarios for preclinical emergencies. 
To gather a wide range of information, we interviewed eleven \glspl{sme} during the \textit{\stageTWO} stage. 

\subsection{Designing new Devices during Online Workshops}

Due to the Covid-19 pandemic, we conducted all stages of this project digitally. 
Corresponding to our situation, Galabo et al. \cite{galabo:redesigning:2020} describe principles of how to plan and execute digital design workshops.
They recommend considering the possible technical limitations and the participants' digital literacy while planning the event.
This consideration helps to choose an appropriate interaction platform, which can also be low-tech (phone and e-mail), if necessary.
Furthermore, the event facilitators should plan short-term activities and provide clear instructions to the participants to reduce information overload.
For example, instructors should provide a program guide to the participants enabling them to check their progress of completing a task.
Preparing a digital whiteboard with predefined design frames, icons, and other elements can help to get started with the design process.
Additionally, ice breaker challenges are important at the beginning of a digital workshop.
They encourage participants to interact with each other and to learn the basic interaction features of the platform.
\section{Our ACTA Approach} \label{sec:project}

\begin{figure*}
    \centering
    \includegraphics[width=2\columnwidth]{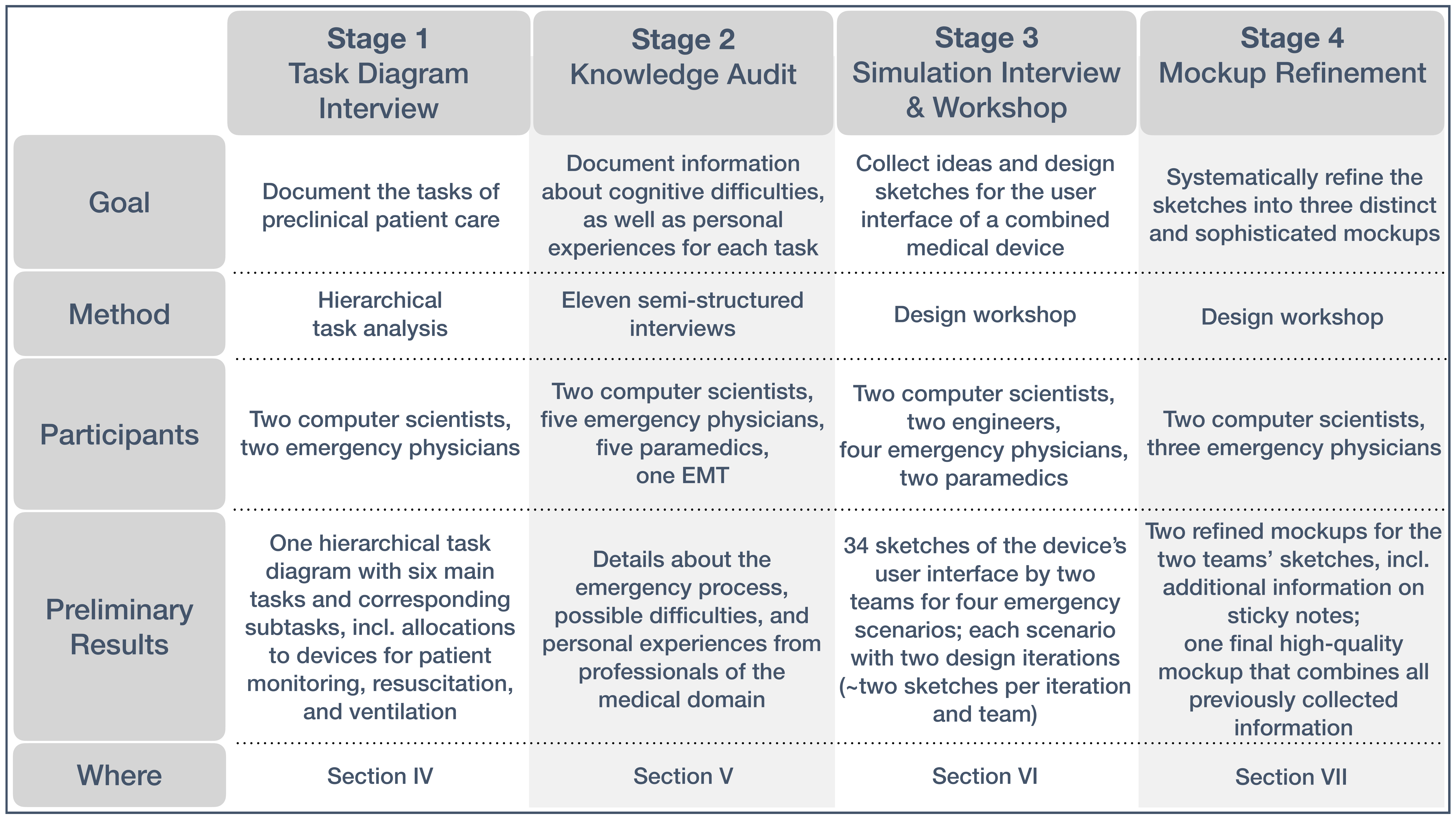}
    \caption{Overview of our customized ACTA approach. All stages were conducted virtually.}
    \label{fig:approach_overview}
\end{figure*}

This section introduces our customized \gls{acta} approach.
We explain each stage of the method on a high level, provide a guide about where to find more details, and present an overview of the participants across the \gls{acta} stages.

Figure \ref{fig:approach_overview} illustrates our overall \gls{acta} approach and emphasizes its four sequential stages represented as the figure's columns. The rows describe each stage's goal, the employed method, the number of participants and their expertise, the preliminary results, and the section detailing the stage.

The authors affiliated with the \gls{uhh} were responsible for the requirements engineering and for designing the new user interface.
The authors from the \gls{uke} brought the expertise in the medical domain and were responsible for evaluating the new device. 
We decided to work with a variation of \gls{acta} since preclinical patient care requires many cognitively demanding tasks in various stressful situations.
Moreover, \gls{acta} enables non-psychologically-trained practitioners to conduct a cognitive analysis of the situations' tasks.
We customized the third and fourth stage of the method since (1) physical simulation interviews were not possible, and (2) our overall aim was to provide a design of the new device's user interface.

\textbf{Stage 1: \stageONE}.
In this stage, we aimed to get an overview of the tasks across all roles in an emergency team during preclinical patient care.
We identified the tasks through two semi-structured interviews with the paper's authors from the \gls{uke} following the \gls{acta} guidelines by Militello and Hutton \cite{militello:ergonomics:1998}.
As a preparation for the interviews, our partners of the \gls{uke} created a \textquote{patient story} document which covers the procedure of most emergency cases.
The output is a \gls{hta} diagram, which serves the next stage as the basis for in-depth interviews with \glspl{sme}. 
Section \ref{sec:stage_1} provides further details for this stage.

\textbf{Stage 2: \stageTWO}.
Based on the previous stage's \gls{hta} diagram, the \stageTWO delves deeper into the analysis of cognitive difficulties of each documented task.
The \stageTWO helps to unveil the most important expertise of the \glspl{sme} for the tasks and discusses personal experiences.
We applied semi-structured video interviews with eleven \glspl{sme} and organized these interviews around knowledge categories like \textquote{diagnosing and predicting}, \textquote{situation awareness}, and \textquote{recognizing anomalies} \cite{militello:ergonomics:1998}.
We detail on the \stageTWO in Section \ref{sec:stage_2}.

\textbf{Stage 3: \stageTHREE}.
This stage is about confronting the participating \glspl{sme} with a challenging scenario that covers the tasks of the \gls{hta} diagram and asking the participants to assess the corresponding situations and actions concerning their cognitive difficulties.
We modified this simulation interview to a digital design workshop that outputs a series of sketches for a combined device's user interface, designed by the \glspl{sme}.
Militello and Hutton suggest using an already existing scenario for this stage \cite{militello:ergonomics:1998}, for example, one that is typically used during medical training.
Therefore, we reused the \textquote{patient story} that the authors of the \gls{uke} created as basis for the \stageONE.
The patient story starts with a scenario that covers only one of the existing medical devices for preclinical patient care.
Following scenarios become more complex and include further devices.
We confronted the participants of the workshop with parts of the patient story and iteratively asked them to create and evaluate user interface sketches for each scenario.
Furthermore, we included results from the previous Knowledge Audits in the workshop and asked detailed questions regarding difficulties in handling the devices and possible technical failures.
Section \ref{sec:stage_3} provides all details for this stage.

\textbf{Stage 4: \stageFOUR}.
In this final stage, we aggregated the results of all previous steps and solidified the collected information and designed sketches into more detailed mockups of the device's user interface. In upcoming phases of the overall industry project, this mockup can be evaluated by further \glspl{sme} and, finally, implemented with code.
To create the mockups, we invited emergency physicians, engineers, and product managers to another digital design workshop.
We describe more details in Section \ref{sec:stage_4}.
\section{Stage 1: \stageONE} \label{sec:stage_1}

\textbf{Description}. 
Before designing a new graphical user interface and alarm management system, we had to understand the domain and the context in which emergency teams use the medical devices for preclinical patient care. 
According to Militello and Hutton, practitioners can elicit a broad overview of tasks and identify demanding cognitive elements during the \stageONE  \cite{militello:ergonomics:1998}.
The authors suggest performing preliminary interviews with \glspl{sme} to highlight three to six steps of a typical scenario.
Once done, the interviewed \glspl{sme} have to divide the steps into tasks and sub-tasks.

\textbf{Participants}. 
Two of the authors from the \gls{uhh}, a postdoctoral requirements engineering researcher and a master's student, interviewed two of the authors from the \gls{uke} who have several years of preclinical patient care experience from hundreds of deployments as emergency physicians.

\textbf{Process}. 
We performed a \glsfirst{hta} \cite{marr:vision:1982, adams:eid:2012} and documented the results of the analysis in an \gls{hta} diagram.
We created the diagram over four iterations, via digitally conducted interviews and asynchronous exchange via email.
The first iteration was an open interview with one of the \glspl{sme} to get a first overview of the steps done in preclinical patient care.
We used the video conference tool \emph{Zoom}\footnote{\url{https://zoom.us}} for this interview since it allowed us to share the content of our computer screen to collaboratively create a first version of the \gls{hta} diagram.
Then, we forwarded the first iteration to the third author from the \gls{uke}, a senior emergency physician, who checked the diagram for completeness and validity.
An idea for the improvement of the diagram was to include an explicit allocation of the tasks to the medical devices that are required for the task's accomplishment.
This allocation helped us during the \stageTWO to better focus our interview questions on the usage of the devices and the corresponding difficulties.
We continued editing the diagram over three additional iterations until all participants found the diagram meaningful and clear enough as a basis for the project's second stage.

\begin{figure*}
    \centering
    \includegraphics[width=2.045\columnwidth]{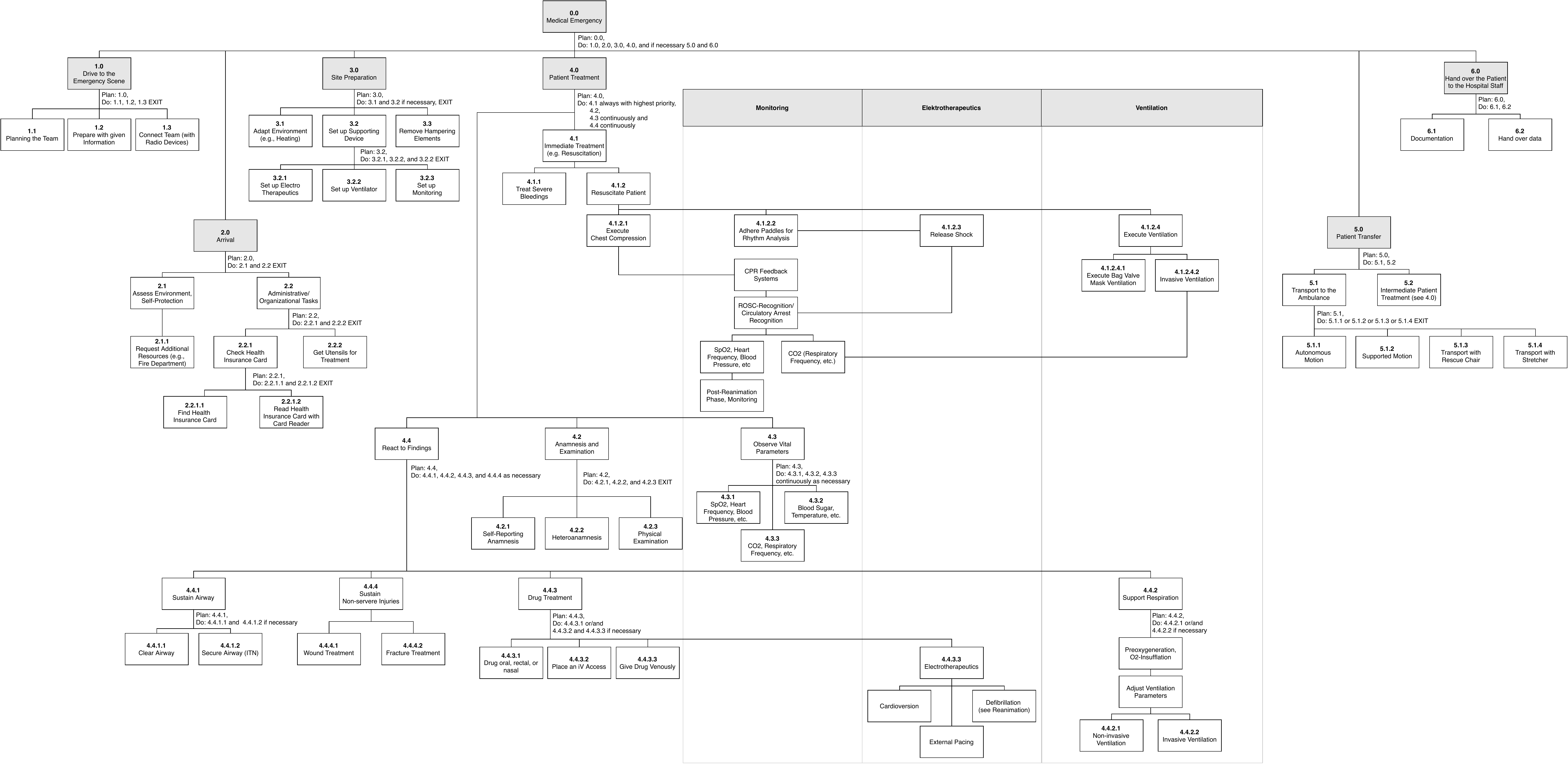}
    \caption{Result of stage 1: A hierarchical task diagram for preclinical patient care (translated from German to English).}
    \label{fig:hta}
\end{figure*}

\textbf{Results}. 
Figure \ref{fig:hta} displays the \gls{hta} diagram and illustrates the hierarchy and order of the emergency-related tasks relevant to the project.
The diagram starts with \emph{task 1.0 -- drive to the emergency scene} and ends with \emph{task 6.0 -- hand over the patient to the hospital staff}.
Following the recommendation of Militello and Hutton \cite{militello:ergonomics:1998}, we restricted the process to six high-level tasks.
For each high-level task, we defined sub-tasks to include further details.
Moreover, the diagram displays the preconditions that have to be in place for the specific task's execution.
For example, an emergency physician can only release a shock via the resuscitation device if the corresponding adhesive pads are attached to the patient's chest.

We modified the diagram's original design by introducing three columns, each belonging to one of the medical devices we want to merge.
By aligning the diagram to our purposes, we gained the following advantages:
\begin{itemize}
    \item The diagram helped us in understanding which tasks require what device. We learned that the medical devices are mostly relevant in \emph{task 4.0 -- patient treatment} while they also affect the necessary efforts to accomplish \emph{task 3.0 -- site preparation} and \emph{task 5.0 -- patient transfer}.
    \item During the \stageTWO phase, we were able to prioritize the tasks for each interview depending on the expertise of the respective interviewee.
\end{itemize}

\begin{center}
\begin{minipage}{\columnwidth}
    \begin{theo}
    A\textbf{Conduct collaborative interviews in multiple iterations}.
    The \gls{acta} guidelines suggest having surface-level interviews for creating the \gls{hta} diagram.
    We found the digital and synchronous interviews via Zoom beneficial to discuss the diagram's notation and content in detail with the \glspl{sme}.
    In particular, sharing the screen content and creating the diagram together during the first iteration helped us to better understand the overall procedure and allowed us to ask questions to the \glspl{sme} in case 
    something was unclear.
    
    We further extended the interviews with asynchronous exchanges over multiple iterations to give both parties enough time to reflect on the diagram and to ensure a common understanding.
    \end{theo}
\end{minipage}
\end{center}
\section{Stage 2: \stageTWO} \label{sec:stage_2}

\textbf{Description}. 
We conducted semi-structured interviews with emergency physicians, paramedics, and \glspl{emt}.
Our goal was to create a \emph{knowledge audit table} and a preliminary \textit{cognitive demand table} documenting the cognitive challenges corresponding with the \glspl{sme} tasks during preclinical patient care.

\textbf{Participants}. 
Two of the authors with a computer science background interviewed five anesthesiologists that also act as emergency physicians, five paramedics that completed a two year-long medical apprenticeship (\emph{Notfallsanitäter} and \emph{Rettungsassistent}), and one \gls{emt} that completed a three month-long training (\emph{Rettungssanitäter}). 
The anesthesiologists are employees of the \gls{uke}. 
Two of the paramedics and the \gls{emt} currently work as product managers at one of our industry partners.
We also interviewed two employees of Hamburg's fire and rescue services and one paramedic of the \emph{Bundeswehr} (German armed forces).
At the time of the \stageTWO, the interviewees participated in 258 to more than 7000 emergencies and worked in the medical domain for four to 21 years.

\textbf{Process}. 
We conducted an interview with each \gls{sme} via a recorded video conference. 
The interviews were 45 to 60 minutes long.
After introducing ourselves and asking demographic questions, we explained the components of the \gls{hta} diagram.
For each \gls{sme}, we pre-selected a subset of the diagram's high-level tasks, depending on the interviewee's profession.
For each selected task, we asked questions based on the \gls{acta} guidelines \cite{militello:ergonomics:1998} to elicit details about the task's cognitive demands. 
In the following, we summarize the questions as suggested by the guidelines:
\begin{itemize}
    \item \emph{Big picture}: What is the big picture of the task?
    \item \emph{Past and future}: Have you been to situations in which you immediately knew what to do in the next moment?
    \item \emph{Noticing}: Have you been to situations in which something unexpected occurred?
    \item \emph{Equipment difficulties}: Have you been to situations in which you could not rely on the devices, for example, since they displayed misleading information?
    \item \emph{Job smarts}: Do you have tricks to accomplish the task more efficiently?
\end{itemize}
The first question served as a conceptual entry point to the specific task where the interviewee explained its purpose and all its components.
After this, we asked more detailed questions on the cognitive difficulties.
Here, we focused primarily on the devices that the practitioners use during the task.

The \gls{acta} method also defines further questions that ask for possible process anomalies, improvising techniques, and the practitioner's self-monitoring during the task execution. 
We stopped explicitly asking these questions after conducting the first three interviews because they have a high overlap with the other questions.
Moreover, they did not help in eliciting further information that is related to the applied devices.

After the interview, we transcribed the interviewee's statements and categorized them in the knowledge audit table and summarized our findings in the preliminary cognitive demand table. 
Both tables act as a basis for the workshop of stage 3.

\textbf{Results}.
\begin{figure*}
    \centering
    \includegraphics[width=2.045\columnwidth]{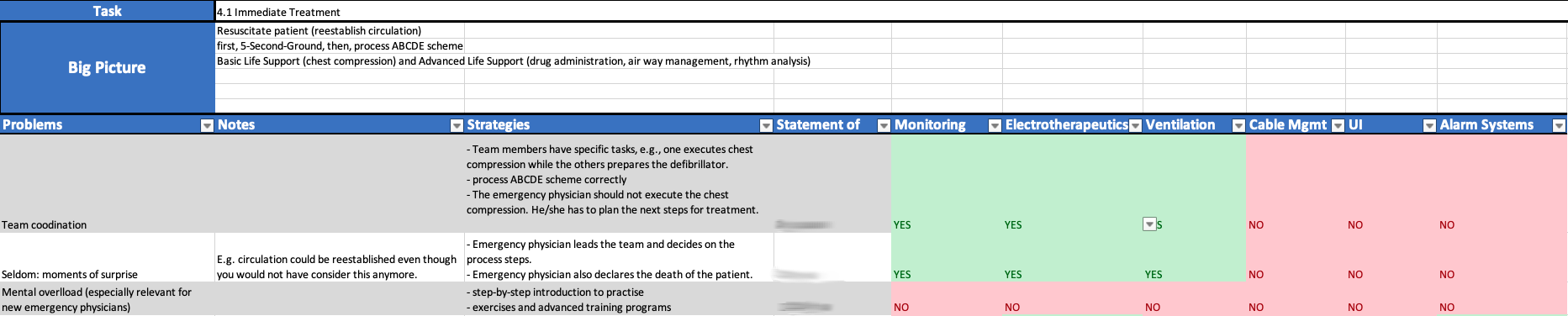}
    \caption{Result of stage 2: A snippet of the knowledge audit table.}
    \label{fig:knowledge_audit_table}
\end{figure*}
Figure \ref{fig:knowledge_audit_table} displays a small part of the knowledge audit table.
We created one spreadsheet page for each high-level task (1.0, 2.0, ...).
The sheets contain the task's name and its 'big picture' on the top. 
We created a row for every statement of the interviewed \glspl{sme}.
For each statement, we documented the addressed \emph{problem}, \emph{cues}, \emph{strategies}, and an ID of the corresponding \gls{sme} to keep track of the participants expertise, role, and experience. 
We further used boolean fields to indicate whether the statement belongs to the devices for \emph{monitoring}, \emph{electrotherapeutics}, and/or \emph{ventilation} and to the categories \emph{cable management}, \emph{graphical user interface}, and/or \emph{alarm systems}.

After conducting and evaluating each interview, we collected the cognitive demands that at least two interviewees mentioned in a cognitive demand table to highlight their importance for the upcoming workshop.
For example, nine participants stated that specific measurement methods to monitor a patient's condition are not reliable in cold environments.
The dissonant measurement data, which the monitoring device shows, could confuse inexperienced practitioners and lead to unhelpful treatment methods. 
Moreover, eight interviewees stated that the devices' placement is often difficult on the emergency scene due to limited spatial conditions, e.g., in narrow bathrooms or piled-up living rooms. 
Hence, the devices must be small in size and number and flexible in their placement possibilities.
\begin{center}
\begin{minipage}{\columnwidth}
     \begin{theo}
     C\textbf{Transcribe and evaluate each interview during the \stageTWO stage}.
     Getting to know the terms and processes of a specialized domain is not trivial. 
     A lack of detailed knowledge can lead to imprecise interview questions and missing follow-up questions.
     Since we wanted to use the interviews' results primarily as a knowledge base for the workshop, we had to understand the emergency-related domain, including edge cases.
     
     Therefore, we transcribed each interview and highlighted statements that we did not fully understand yet.
     In upcoming interviews, we asked the participants particularly about these statements to improve our understanding of the domain.
     \end{theo}
\end{minipage}
\end{center}
\section{Stage 3: \stageTHREE} \label{sec:stage_3}

\textbf{Description}.
The purpose of the Simulation Interview is to confront \glspl{sme} with challenging scenarios to understand their decisions in specific contexts on a deeper level.
The context dimensions that characterize our case are, for example, the place (outside on a street, in a car, fifth floor, etc.), the weather (sunny, dark, rainy, etc.), and the patient's condition.

In contrast to the original \gls{acta} method, we not only required insights in the problem space but also information about the solution space to reach our project's goal -- a design for the graphical user interface of a combined medical device.
To achieve this, we designed a full-day digital creativity workshop that should achieve creative yet practical sketches for a potential user interface.

\textbf{Participants}. 
Ten people with different expertise participated in the workshop.
Among the participants were two of the authors with background in computer science and requirements engineering, two paramedics, four emergency physicians, and two engineers. 
The paramedics and emergency physicians all have several years of experience in preclinical patient care and used multiple emergency-related devices during their careers.
The two engineers are responsible for the user interface development for the medical devices of one of our industry partners.
We, the requirements engineers, acted as the workshop's moderators and were responsible, among other, for including the second stage results in this workshop.

\textbf{Process}. 
\begin{figure}
     \includegraphics[width=\columnwidth]{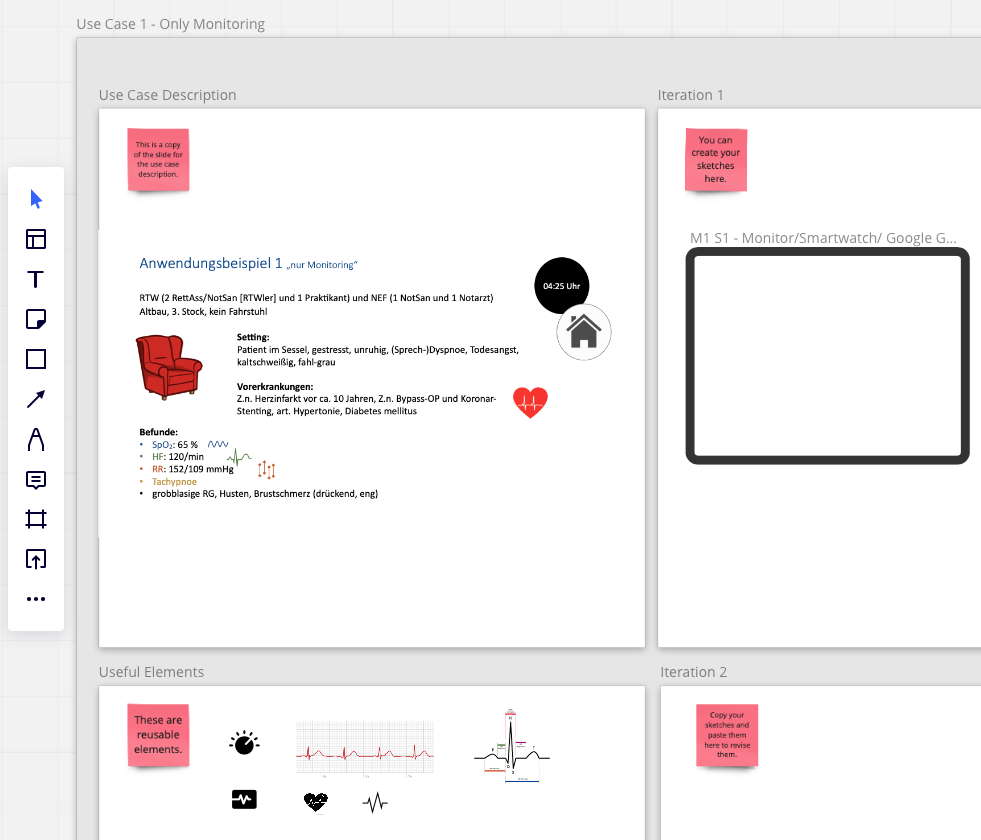}
     \caption{Workshop preparation using Miro's digital whiteboard.}
     \label{fig:miro}
\end{figure}
We conducted the digital workshop using the online collaborative whiteboard platform \emph{Miro}\footnote{\url{https://miro.com}} and Zoom.
On Miro, teams can work together in real-time by, for example, creating sticky notes, using wireframes, inserting images, text, shapes, and more. With Zoom, teams can talk to each other and use the screen-sharing feature for moderation purposes.

In the workshop, we had two cross-functional teams working on their design ideas together.
As explained in Section \ref{sec:project}, the authors from the \gls{uke} created a \textquote{patient story} describing a challenging and realistic emergency case that gradually includes more medical devices.
We used this patient story to structure the workshop and performed the following steps iteratively for all four use cases of the story:
\begin{enumerate}
    \item We present the respective use case to the participants.
    \item The participants design sketches of a user interface for the use case.
    \item The teams present their solutions and provide feedback.
    \item The teams improve their sketches based on the feedback and the impression of the other team's solution.
\end{enumerate}
During the workshop planning, we first considered no presentation and feedback steps to achieve more diverse and potentially more creative ideas. 
However, we also needed practical results to progress in the project faster.
In the way described above, the teams could either ignore the other team's feedback and results or include it in their sketches during the improvement step.

Figure \ref{fig:miro} shows a snippet from the Miro board we prepared for the workshop.
Each board has four grey rows each containing two white rows -- the figure shows one of the grey rows.
Every grey row represents one use case.
The upper white row is for the first round of sketches of the workshop step (2), while the bottom white row is for the improvement iteration in step (4) after the feedback round.
We included the slides summarizing the patient story's respective use case on the left side of the upper row to support the participants during the design process.
The bottom row's left side included frequently used elements, like a heart rate graph, that can be hard to discover with Miro's features.
Each row's right side provides enough space for sketches and notes.

Since we wanted the \glspl{sme} to ideate and consider options like multiple monitors for a single device or alternative devices like smartwatches or smart glasses, we introduced a notion to the sketches.
On the upper right in Figure \ref{fig:miro}, there is an empty frame that has the title \textquote{M1 S1 -- Monitor}.
The \textquote{M} refers to the monitors of the hardware devices.
For example, if the participants consider one emergency team member with a smartwatch and the other members with a conventional monitoring device, they have to design the user interfaces of two hardware devices.
Therefore, they could name the smartwatch \textquote{M1} and the other device \textquote{M2}.
\textquote{S1} refers to the screen.
If a user can open multiple views on the monitor of one device, the design teams could mark these screen variations with \textquote{S1} and \textquote{S2} respectively.
Additionally, the participants should specify with a short description whether the hardware device is a classical monitor, a smartwatch, a smart glass, or any other device.
With this notion, we could get an idea of the navigation flow and track changes between the iterations and the use cases of the patient story.

For documenting the teams' mutual feedback, we also used a feature of Miro.
Miro allows writing board-wide notes that the team can see no matter where they are on the board and how much they zoom in or out.
We documented the feedback there so that the teams can focus on the discussion of their results without having to worry about forgetting any suggestions for improvement.
One of the authors moderated the feedback sessions while another one noted the feedback.
We separated the feedback per use case and used the introduced notation (for example, \textquote{M1 S1 - Smart Watch}) to articulate to which screen the feedback belongs to.

\textbf{Results}.
While we cannot share the exact results due to confidentially reasons, we can report on the workshop results from a process perspective and how the participants perceived the workshop based on their answers in a follow-up survey.

The workshop's total length was six hours, including the introduction, breaks, and the farewell.
In total, the participants created 34 sketches of the screens.
The teams accomplished to create iteration-based sketches with increasing complexity.
Furthermore, the discussions of the results led to new requirements for a combined medical device.
We are pleased with the trade-off decisions that we made during the planning of the workshop as we found the results both innovative yet practical and helpful for further proceedings. 
In the following, we discuss some of these trade-off decisions.

First, we thought about not letting the teams present their results until the very end of the workshop, which might have led to more creative ideas.
However, due to introducing the presentation and feedback session after every use case, the teams had interesting discussions about potential ideas and often found inspiration from the others' ideas.

Second, we thought about having every person working on their own sketches and conducting feedback iterations only within the teams.
Then, the teams would decide on one solution that they would present to the other team at the end of the workshop.
Although we applied this approach successfully in on-site workshops in previous projects, we decided against this approach since the current emergency-related devices offer features that only emergency physicians usually use. 
Paramedics might not have the permission and the training to use these features.
However, our workshop participants have different experiences and diverse vocational backgrounds.
Therefore, we allocated them into teams to have diverse experiences, roles, and backgrounds per team and to let them work together on one Miro whiteboard.

The third trade-off concerns the digital execution of the workshop.
Due to the Covid-19 pandemic, during which all reported steps were conducted, we had to perform digital workshops.
Hence, the participants could stay at home and did not have to commute to a specific venue. 
Yet, it was very difficult to recruit medical \glspl{sme} for a full-day workshop since they were also needed in the hospitals.
Therefore, we also discussed an asynchronous workshop that would have been similar to the DELPHI method \cite{linstone1975delphi}.
This alternative would not require the participants to reserve a full workday but to contribute over several weeks via take-home tasks.
In the end, we chose the one-day workshop since we wanted to avoid that some participants might drop out of the asynchronous workshop due to other obligations.

Our follow-up survey contained multiple five-point Likert Scale questions ranging from very easy (1) to very difficult (5).
Six out of eight workshop participants answered the survey. 
Besides one person that rated the ease of use of Zoom with 4/5, all other participants found the tools very easy to use and gave a 1/5.
Regarding Miro, the answers spread more across 2/5 and 4/5, with four participants giving 2/5.
Five participants rated the workshop as exhausting between 3/5 and 4/5, while 5 means very exhausting.
Only one person outside the moderation team already had experience in sketching user interfaces.
The participants stated that they felt well informed during the workshop and understood what we requested from them.
Nonetheless, two-third describe that they would have preferred an on-site workshop.
All agree that they would not have liked to participate in an asynchronous digital workshop, which includes take-home tasks.

\begin{center}
\begin{minipage}{\columnwidth}
     \begin{theo}
     D\textbf{Make the participants feel welcomed}.
     In a digital workshop, it is challenging to create the working spirit and team dynamics as in a physical face-to-face workshop since the participants have limited interaction options. 

     To reduce these limitations, we started with an introduction round and conducted a ice breaker challenge where the participants had to create a sticky note with their names in Miro. 
     Although these measures hardly compensate for coffee talks or personal interactions, we observed that the overall atmosphere was much better and interactive. Participants got involved in discussions early on and that everyone gave inputs during the design sessions.
     \end{theo}
\end{minipage}
\end{center}

\begin{center}
\begin{minipage}{\columnwidth}
     \begin{theo}
     E\textbf{Participants should not think about the requirements and design process and instead focus on the creative tasks}.
     We wanted to avoid that participants have too many questions about the process and tools, like \textquote{How do I use this tool?}, or \textquote{How can we present our results and give feedback?}.
    
     To let the participants focus on the creative part of the workshop, we prepared the following:
     \begin{itemize}
         \item We assembled the teams beforehand.
         \item We used \textit{breakout rooms} in Zoom to automatically split the participants into teams.
         \item We prepared Miro boards including references to the tasks and spaces dedicated for the sketches.
         \item We conducted short tutorials for Miro and Zoom.
         \item One of us was the main moderator for all feedback sessions.
         This person shared their screen to display the results of the teams and requested the participants to explain their results.
         A second moderator was responsible for noting down all feedback so that the \glspl{sme} can focus on participating in the discussion.
     \end{itemize}
     \end{theo}
\end{minipage}
\end{center}

\begin{center}
\begin{minipage}{\columnwidth}
     \begin{theo}
     F\textbf{Despite our concerns about the limitations of digital workshops, using them led to opportunities we would not have reached otherwise}.
     Certain team dynamics only appear if working in-person, like having distinct chats within teams without the others need to stop talking or quickly creating sketches with a pen.
     Nevertheless, the digital workshop led to the following advantages:
     \begin{itemize}
         \item Focused discussions in which everyone listens to every idea and opinion.
         \item Readable and organized notes with thoughts, ideas, and further information describing the sketches.
         \item Digital sketches that can be copied and edited during further iterations.
     \end{itemize}
     \end{theo}
\end{minipage}
\end{center}

\section{Stage 4: \stageFOUR} \label{sec:stage_4}

\textbf{Description}. 
We refined the sketches of the previous workshop into three sophisticated mockups: one mockup for each of the two teams and a final version that includes the best design ideas of both mockups, according to our assessment.
We conducted a second design workshop to reach this goal.

\textbf{Participants}. 
Two authors of the \gls{uhh}, all authors of the \gls{uke}, and a project manager of one of our industry partners.

\textbf{Process}. 
Like the previous workshop, we realized this phase entirely online via Zoom and Miro. 
We prepared a new Miro board by consolidating and categorizing previously collected knowledge.
For this, we divided the board into three columns. 
On top of the first two columns, we collected the statements of our preliminary cognitive demand table and the notes of the respective team as sticky notes.
We used different colored notes to categorize the statements according to their relevance to the user interface layout, to the monitor (including hardware buttons and size of the monitor), to further hardware of the device (e.g., cables), and to additional devices.
Per column, we placed the respective team's sketches in a row-like structure below sticky notes.
One row referred to one use case of the patient story. 
We added an additional row at the bottom for the final version of each teams's mockup.
The third column was dedicated to the third mockup.
This preparation helped us to get an overview of each team's design ideas and the user interface requirements of the combined medical device.

The workshop took place on two days, each about eight hours long. 
We started day one by introducing the new sketch board to the participants.
Evaluating the ideas of the first team and designing mockups from their sketches took us almost the entire first day.
We realized that the previous workshop produced sketches fast and in good quality, however, with an incomplete scope of functions. 
Due to the limited time per use case, the team focused on the most important features and near-ideal usage situations. 
The sketches did not cover, for example, all parameters for the ventilation or a view for additional settings.
Hence, we had to design and add these mandatory features. 
Nevertheless, we accomplished the second team's mockup and the third and final version on day two since all participants operated well together and the complex use cases were already discussed on the first day.

\textbf{Results}. 
The combined medical device has to cover many parameters. 
We managed to include 21 numerical parameters and additional graphs for ten of the parameters in the screen of our final mockup. 
This screen should have a size of at least twelve inches.
We decided early that users interact with the device via a touchscreen to assure fast access to each parameter.
Yet, it was important to the participating emergency physicians to have a fallback handling via a rotatable and clickable hardware button for situations in which touchscreens do not work well like in rainy weather.
The parameters are separated into three functionally cohesive areas.
These areas are spatially divided from each other via separation lines to assure that a user can quickly retrieve a specific parameter's location.
To avoid information overload on the screen a parameter is only displayed when it is currently measured.

Based on the results of this workshop, further tasks of the overall industry project can be accomplished.
\begin{center}
\begin{minipage}{\columnwidth}
     \begin{theo}
     G\textbf{Have preliminary requirements at hand}. Incomplete or too complex feature sets can prolong the design phase and might lead to outcomes of poor quality.
     
     Therefore, collect and categorize all important information and findings of the previous phases. Check the completeness of the previous workshop results and highlight aspects that were not taken into account yet.
     \end{theo}
\end{minipage}
\end{center}
\section{Discussion} \label{sec:discussion}

\subsection{Applicability of \gls{acta} in this project}

The purpose of the original \gls{acta} method is to elicit the critical cognitive demands on \glspl{sme} during a complex scenario \cite{militello:ergonomics:1998}.
Hence, \gls{acta} only concerns the problem space.
Practitioners that apply this approach aim to get insights into the existing processes of the considered scenario.
These insights are the basis for the upcoming work in the solution space to provide improved processes.
Even though \gls{acta} is more streamlined than \gls{cta}, it includes many extensive steps and artifacts.
Furthermore, it does not focus on the interaction with the applied hardware and software but is a process-level analysis of cognitive difficulties.
On the contrary, we aimed to analyze emergency scenarios with a focus on the usage of medical devices and the difficulties that occur during their usage.
Hence, we -- as requirements engineers -- had to understand the tasks of the given scenario superficially.
However, more importantly, we also had to collect detailed information concerning the hardware and the user interface of the existing devices to design a user interface concept of the combined device.

Despite the gap between the features of \gls{acta} and our goals, we chose this approach since it guided us in structuring the components of emergency scenarios and eliciting cognitive difficulties of specialized tasks.
Moreover, the modular \gls{acta} framework allowed us to adjust single steps. 
For example, we included a design workshop in the Simulation Interview stage.

\subsection{Strengths and weaknesses of the \gls{acta} stages}

We discuss the strengths and weaknesses of the original \gls{acta} stages as we experienced  during the project.

\textbf{Task Diagram Interview}.
The \gls{hta} diagram provided us an overview of a comprehensive emergency scenario.
Three authors designed the diagram during multiple iterations to combine our knowledge in requirements engineering with medical expertise. 
We included an additional table in the diagram to structure the sub-tasks of \textit{task 4.0 - patient treatment} into device-related columns.
We chose this task since it includes the usage of all devices that we consider in this project.
Nevertheless, also tasks 3.0, 5.0, and 6.0 include difficulties concerning the devices' setup, transport, and cleaning (see Figure \ref{fig:hta}).
Hence, it would have been beneficial to us if the \gls{hta} diagram naturally offers a possibility to allocate single tasks to devices that are used to accomplish the task.

On the contrary, tasks 1.0 and 2.0 do not include sub-tasks that are relevant to our considered medical devices.
Although the tasks were not required for the \gls{acta} stages, including details of these steps in the diagram did help the non-medical authors in understanding the domain. 

\textbf{Knowledge Audit}.
The Knowledge Audit was developed to capture information about the expertise of \glspl{sme} while facilitating the corresponding data collection process \cite{militello:ergonomics:1998}. 
We, two of the requirements engineers, conducted eleven interviews with \glspl{sme} of the medical domain.
Per interview, one of us took the interviewer role while the other one took an observer role to document the interviewee's statements.
The original Knowledge Audit provides six basic and two optional probes, each describing different knowledge categories, like the \textit{Big Picture}, or \textit{Job Smarts} \cite{militello:ergonomics:1998}.
We stopped explicitly asking questions of some of the categories after finishing the first three interviews since the interviewees' answers were highly overlapping with the answers of other questions.
For example, we asked the \glspl{sme} if they noticed unusual signs during emergencies that others did not catch (category \textit{Noticing}).
The answers to this question often concerned the environment of the emergency scene or a medical device that was not working as intended. 
Later, we asked questions for \textit{Opportunities/Improvising} and \textit{Equipment Difficulties} and the \glspl{sme} gave us very similar answers to these categories.
Using only a subset of the original categories enabled us to have a more natural conversation with the interviewees and to discuss a higher number of tasks during the time-limited interviews.

Moreover, we realized after the first interviews that the answers to the given probes mostly concern the difficulties of an emergency on a process level but rarely give details about the applied devices.
As a countermeasure, we especially focused on the \textit{Equipment Difficulties} in later audits and asked at the end of the interviews what benefits and drawbacks the \glspl{sme} expect from the combined medical device.

Furthermore, we swapped the roles of interviewer and observer throughout the Knowledge Audit stage to mitigate a bias caused by an interviewer's way of leading the conversation.

\textbf{Simulation Interview}.
\Gls{acta} includes a workshop-like Simulation Interview where participants can physically demonstrate the cognitive and pragmatical difficulties of a specific scenario.
Due to the Covid-19 pandemic, we relied on an entirely digital variant of this workshop.
Moreover, it was not our primary aim to further analyze the problem space but to design first user interface prototypes together with \glspl{sme}.
It was beneficial for the participants that we started with a relatively simple use case but also gave them additional time in the first iteration to get to know the features of Miro.
Moreover, we prepared the Miro design boards with specific design areas for each workshop phase and added specific design elements, like a heart rate graph, to the boards.
Hence, participants did not have to spend time searching for these specific elements online.
All the measures were crucial to the workshop's success since the majority of \glspl{sme} did not take part in a design workshop yet.

According to the answers to the survey that we conducted afterward, the participants did not feel overwhelmed with the digital environment and the design tasks.
Moreover, Miro enabled us to easily copy the designed user interface sketches for improvements in the upcoming iterations and to add and structure additional information on digital sticky notes.
Nevertheless, the survey participants perceived the workshop as demanding and two-thirds would have preferred an in-person workshop.
Reasons for this can be, among others, the missing informal chats during the breaks and the physical design process of in-person workshops that can be perceived as more interactive than the digital alternative.

\textbf{Cognitive Demand Table}.
Due to our exploration in the solution space during the \stageTHREE stage, we did not rely on a final version of the cognitive demand table.
Nevertheless, this table, which represents a summary of all cognitive difficulties of the emergency scenario, can be a useful source for further steps of the overall project.
For example, the hardware engineers will require information about the \glspl{sme} preferences concerning the device's size or the arrangement of cable ports.
Therefore, we conclude that the cognitive demand table is a useful artifact to compactly document all findings of the previous stages.
However, depending on the usage of \gls{acta} in specific projects, it might not always be required.
\section{Conclusion} \label{sec:conclusion}

We presented our six lessons learned from customizing and applying the \glsfirst{acta} method to design the user interface of a novel, sophisticated medical device for preclinical patient care.
We separated our approach into four stages.

In the \textit{\stageONE} stage, the first two authors, who are requirements engineering experts from the \gls{uhh}, iteratively performed interviews and exchanged information with two doctors of the \gls{uke}, who are experts in the emergency medical care domain, to create a hierarchical task diagram.
This diagram illustrates the tasks of preclinical patient care from when the emergency team is informed about the case until handing-over the patient to the hospital.

During the second stage, the \textit{\stageTWO}, both the requirements engineering and the domain experts performed semi-structured interviews with eleven \glspl{sme} in preclinical patient care.
Our goal in this stage was to better understand the cognitive challenges that emergency teams face during each task of the previously created diagram.

In the third stage, \textit{\stageTHREE}, we conducted a design workshop together with eight other participants with expertise in emergency medicine and engineering.
We split the participants into two cross-functional teams, which created 34 sketches for the user interface of the envisioned medical device.
We included the interview results of the previous stage to create awareness for specific issues in handling existing devices.

During the last stage, \textit{the \stageFOUR}, we used the sketches of both teams to develop two more sophisticated mockups. 
Furthermore, we created a final mockup that combines all previously collected knowledge.

The entire customized \gls{acta} process helped us to achieve innovative yet practical results.
The project consortium jointly agreed that the approach was a major enabler toward achieving a feasible combined medical device that will support the emergency teams in preclinical patient care.

\section*{Acknowledgement}

This work has been partially funded by the German Federal Ministry of Education and Research (grant number 13GW0333D VentCore). The responsibility for the content of this publication rests with the authors.

\bibliographystyle{abbrv}
\bibliography{lib}

\end{document}